\documentclass[pra,twocolumn,showpacs,preprintnumbers,amsmath,amssymb,floatfix]{revtex4}

\usepackage{graphicx}
\usepackage{dcolumn}
\usepackage[squaren]{SIunits}

\def\fig#1{Fig.~\ref{#1}}

\def\sec#1{Sect.~\ref{#1}}
\def\Sec#1{Section~\ref{#1}}
\def\eq#1{Eq.~(\ref{#1})}
\def\fig#1{Fig.~\ref{#1}}
\def\Fig#1{Figure~\ref{#1}}

\def\be#1{\begin{equation}\label{#1}}
\def\ee{\end{equation}}
\renewcommand{\vec}{\mathbf}

\begin{document}
\title{Clusters under strong VUV pulses: \\
  A quantum-classical hybrid-description  incorporating plasma effects}
\author{Ionu\c{t} Georgescu, Ulf Saalmann and Jan M. Rost}
\affiliation{Max Planck Institute for the 
  Physics of Complex Systems\\
  N\"othnitzer Stra{\ss}e 38, 01187 Dresden, Germany}
\date{\today}

\begin{abstract}\noindent 
  The quantum-classical hybrid-description of rare-gas clusters
  interacting with intense light pulses which we have developed
  is described in detail.  
  Much emphasis is put on the treatment of screening electrons
  in the cluster which set the time scale for the evolution of
  the system and form the link between electrons strongly bound
  to ions and quasi-free plasma electrons in the cluster. 
  As an example we discuss the dynamics of an Ar$_{147}$ cluster
  exposed to a short VUV laser pulse of 20\,eV photon energy.
\end{abstract}

\pacs{36.40.Gk,         
  32.80.-t,             
  52.50.Jm,             
  36.40.Wa              
} 

\maketitle

\section{Introduction}
The interaction of rare-gas clusters with intense laser pulses
has attracted considerable attention over the last decade since 
spectacular experiments showed how effective energy from the
laser beam can be coupled into the cluster, culminating in the
demonstration of fusion in exploding deuterium clusters
\cite{zwsm+00}. 
Other efforts aim at understanding the interaction of matter beyond a
few atoms with light of different wavelength, motivated by the fact
that little is known about such processes with intense light of
frequencies different than the infrared (IR) range around 800 nm
wavelength.  First experiments at FLASH in Hamburg have revealed very
efficient energy absorption at 98 nm wavelength \cite{wabi+02} and
have triggered theoretical research
\cite{sagr03,siro04,jura+05,wasa+06} as recently reviewed
\cite{sasi+06}. 

Clearly, the time-dependent description of a finite many-body system
interacting with a laser pulse is challenging.  A full quantum
description is numerically impossible and would probably provide
little insight conceptually.  One can observe two different strategies
in the literature to tackle this problem: (i) A more or less full
quantum approach for a single atom in the cluster treating the
influence of the other atoms and ions in the cluster more
approximately as a kind of environment \cite{dero+06, sagr03}.  (ii)
A classical propagation of all charged particles in time (ions and
electrons) with quantum mechanical elements added in form of rates on
various levels of sophistication
\cite{isbl00,lajo00,newo+00,siro02,saro0203,juge+04,jufa+04,foza+05}.

We have followed the latter route, describing the motion of all
charged particles classically, while they interact and are subjected
to the external dipole-coupled laser field.  Initial bound electronic
motion is not treated explicitely, but only in form of (quantum)
ionization rates which are integrated into the classical dynamics with
a Monte Carlo scheme, see \cite{sasi+06}.  The approach works very
well, as long as the electrons, once ionized from their mother-ions, 
behave essentially classically.  This is the case if they are subject
to a strong external field (as it is the case with strong IR pulses)
or are instantaneously free (as in the case of hard X-rays).  Under VUV
radiation, however, the photo-electrons stay often very close to their
mother-ions or other ions, effectively screening them and modifying
subsequent ionization processes.

The reason for this behavior is the size of the quiver motion
(typically less than a ground state electron orbit of the rare-gas
atom) and the small kinetic energy which remains of the photon
energy in excess of the ionization potential.  The latter is not the
case for hard X-rays where the photon energy is high enough to remove
the photo-electrons completely from the cluster (at least for moderate
cluster sizes).  For intense IR-fields, on the other hand, the photon
energy is much too low for the photo-electrons to leave the cluster
instantaneously, but the quiver motion is (for typical
intensities) of the size of the cluster or even larger implying
that the photo-electrons certainly will not remain in the
vicinity of specific ions.
Rather, they are dragged back and forth through the cluster by
the laser field.  
Hence, the photo-ionization of ions surrounded by screening
electrons is a phenomenon unique to VUV radiation, adding to the
challenge for a theoretical description.  
On the other hand, as will become clear
subsequently, exactly those electrons which screen individual ions define a
timescale, suitable to formulate a coarse grained electron dynamics in
the cluster. It is the key to incorporate physical processes which
in our approach lie at the interface of classical and quantum
description, such as the influence of the surrounding charged 
particles (ions and electrons) on the photo-ionization rate of an ion
in the cluster.

In the next section we summarize our quantum-classical hybrid
description which contains the quantum photo-ionization rates of
multi-electron ions, the classical propagation, and, as an element in
between, the treatment and identification of those electrons which
screen the ions. In \sec{sec:plasma} we explain how to deal 
with photo-ionization of bound electrons into the plasma of cluster 
electrons. \Sec{sec:run} discusses as an application and 
illustrative example the 
illumination of  Ar$_{147}$ with a VUV pulse of
\unit{62}{\nano\meter} wavelength. The paper ends  
with a short summary in \sec{sec:sum}.

\section{Hybrid-description of time-dependent cluster evolution}
\label{sec:hybrid}
The interaction of a cluster with intense radiation can be partitioned
into three parts: (A) atomic ionization, (B) cooperative interaction
of ions among each other and with electrons, and (C) relaxation.
During phase (A) which lasts approximately until every second atom in
the cluster is (singly) ionized, one can simply apply the atomic photo
ionization cross sections for many-electron atoms supplied, e.g., in
\cite{co81,chco+92}.  Gradually, the bound electrons feel the cluster
environment, which has roughly three effects (i) the ionization
potential is lowered through close by ions, (ii) previously ionized
electrons trapped by the cluster screen the ion under consideration,
(iii) the global field, generated by electrons and ions, modifies the
ionization potential as well.  We will treat all three effects which
happen during phase (B) of the dynamics in \sec{sec:plasma}.
Here, we describe briefly how to extrapolate the known photo-ionization
cross sections below the ionization threshold.

\subsection{Atomic photo-ionization}
\label{sec:photo}
The calculation of photo-ionization and excitation for individual
energy levels of isolated atoms and ions is straight forward.  Including the
(quantum) photo rates into the classical propagation of charged
particles in the cluster, we do not resolve the angular dependences of
the photo cross section.  Averaging over angular degrees of freedom
considerably simplifies the cross sections which only depend on the
radial wavefunctions for the respective mean initial and final
energies of the photo transition, for a similar philosophy see
\cite{wasa+06}.  We start from the dipole matrix element for
linearly polarized light along the $\hat z$ direction and consider the
transition between initial and final atomic states with well defined
orbital quantum numbers $lm$
\be{dipolematrix}
\langle f|z|i\rangle = d_{r}(\ell_{f},\ell_{i}) \int d\Omega\,
Y^{*}_{\ell_{f}m_{f}}(\Omega)\cos\theta\, Y_{\ell_{i}m_{i}}(\Omega)
\ee
with the radial dipole matrix element
\be{radialdipole}
d_{r}(\ell_{i},\ell_{f}) =\int_{0}^{\infty}dr 
u_{\ell_{f}}(r)\,r\,u_{\ell_{i}}(r)\,.
\ee
Within the independent particle picture all states in a shell defined
by orbital angular momentum $\ell_{i}$ are degenerate and have the
same radial wavefunction $u_{\ell_{i}}(r)$.  It is an eigenfunction to
a spherically symmetric effective single particle potential obtained
(along with the eigenfunction) from the Cowan code \cite{co81}.
Within this mean field approximation the full photo cross section is
easily obtained by summing over all available final states and
averaging over the initial states in the shell $\ell$,
\be{photoabsorb}
\sigma_{\ell}(\omega)=\frac{w_{i}}{3}\frac{4\pi^{2}\alpha\omega}
{2\ell+1}[(\ell+1)d_{r}^{2}(\ell,\ell+1)+
\ell d_{r}^{2}(\ell,\ell-1)]\,.
\ee
Here, $w_{i}$ gives the number of available initial 
electrons.

Furthermore, in the cluster the ions are surrounded by other electrons
and ions.  The latter lower the potential barriers for ionization into
the cluster.  Hence, what is discrete photo excitation in an isolated
atom becomes at the same photon energy ionization into the cluster.
Therefore, we need an interpolation of the discrete excitation
spectrum distributing the oscillator strength over the respective
photon energy interval.

\begin{figure}[tbp]
\centering
    \includegraphics[width=0.9\columnwidth]{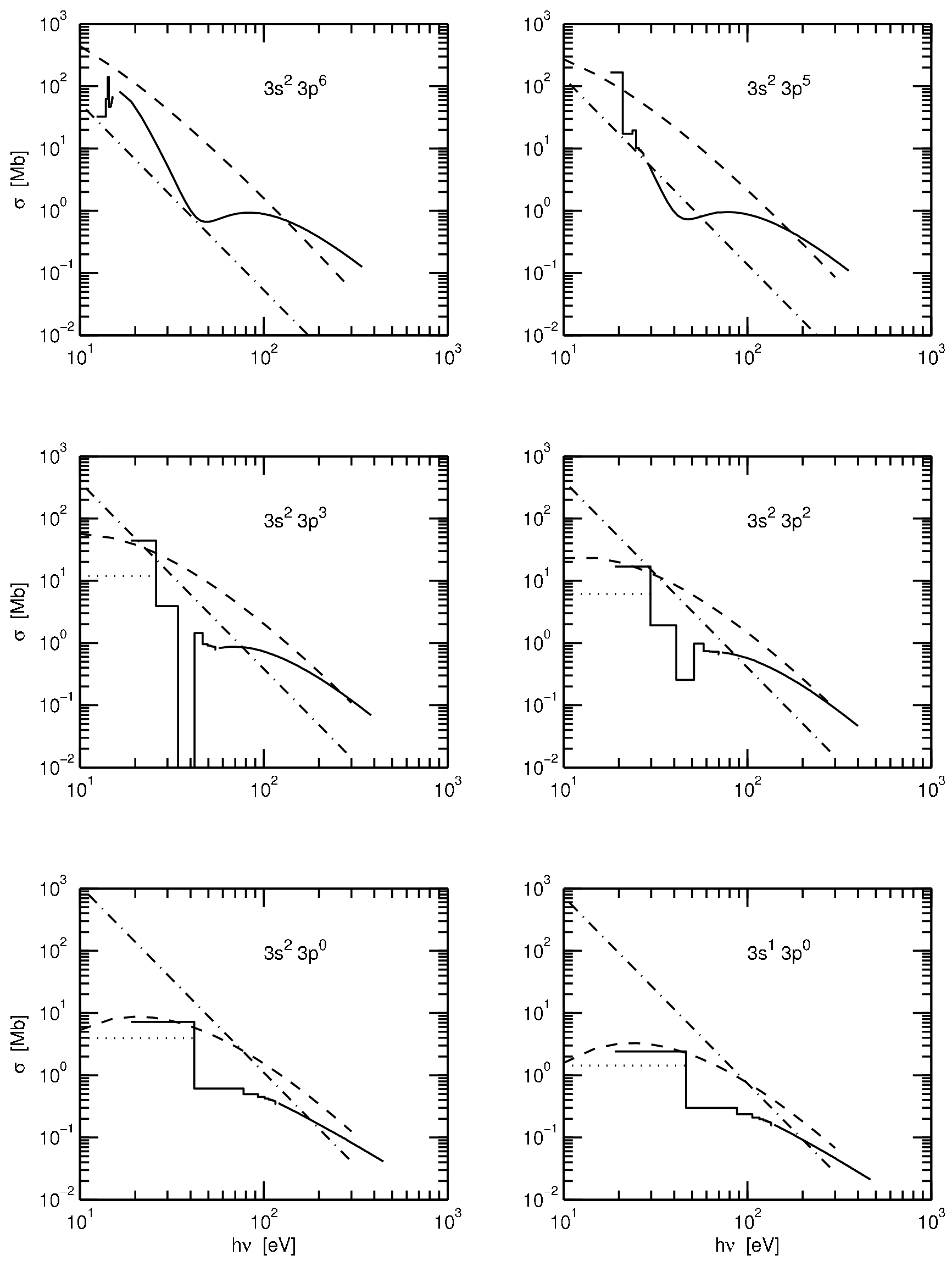}
    \caption{Atomic photo-ionization rates extrapolated below the
    threshold according to \cite{ro95} (dashed) and by interpolation 
    of the discrete spectrum (solid). The dashed-dotted line is an 
    approximation with hydrogenic wavefunctions, see text. The 
    initial configuration is displayed in each panel.}
    \label{fig:photorates}
\end{figure}

In the following, we compare two different interpolations.
The first one is taken from \cite{ro95}. There, approximate 
analytical expressions for the photo-absorption of many electron atoms 
are derived. The corresponding results for the Ar atom are shown in 
\fig{fig:photorates} with dashed lines.  For comparison, the 
 cross sections with hydrogenic wavefunctions
 \cite{zera66} are also shown. 
 In the second approximation we define a continuous photo-absorption in 
 the region of discrete spectral lines by demanding that the renormalized photo 
 excitation cross section merges smoothly with the photo-ionization 
 cross section at threshold \cite{fr98}.
 This is achieved by distributing the 
 the oscillator  strength $f_{n}$ of a spectral line $E_{n}$ over 
 an interval half  way to each of the  adjacent lines
 such that 
 \be{dfde}
  \sigma_{n}(\omega)=\frac{2\pi\alpha f_{n}}{(E_{n+1}-E_{n-1})/2}\,,
 \ee
 where $\sigma_{n}(\omega)$ is now the interpolated 
 photo-absorption cross section  \eq{photoabsorb} for $(E_{n-1}+E_{n})/2<\omega<(E_{n}+E_{n+1})/2$.
 The result (solid line in \fig{fig:photorates})
shows reasonable agreement with the analytical approximation for low and 
high energies. For intermediate energies, the well known Cooper minima
lead to considerably lower values than in the analytical approximation 
which does not account for this interference effect.
Hence, we will use in the following the approximation \eq{dfde}.

\subsection{Classical propagation for the Coulomb interaction
  under VUV radiation}\label{sec:clpr}
The propagation of the classical particles is in principle straight 
forward. More refined methods, such as tree-codes are only worth the 
effort of coding for large clusters ($10^5$ electrons and ions and 
more). Another issue is the Coulomb interaction.
Using the real Coulomb
potential with its singularity is numerically very costly (small
time steps close to the singularity) and leads for more 
than two bound electrons to artificial autoionization since one
electron can fall into the nucleus (below the quantum ground state
energy) and another one can be ionized with the released energy.
In strong field physics (at IR frequencies) the so called
soft-core potential 
\be{ushape} 
U(r) = - \frac{Z}{(r^{2}+a^{2})^{1/2}} 
\ee
has been used routinely where the singularity is cut off by the
smoothing parameter $a$ chosen to get potential depths
(slightly) below the true ionization potential of the atom or
ion.   
As long as the quiver amplitude $x_\mathrm{quiv}=
F/\omega^{2}\gg a$, the cut off is irrelevant for the dynamics,
as it is typically the case for a strong pulse (for $3\times 
10^{16}\mathrm{W/cm}^{2}$ and 800 nm wavelength, 
$x_\mathrm{quiv} \approx 500$\,a$_0$ while $a$ is of the order
of 1\,a$_0$).  
However, at 10 times higher photon frequency and a factor 100
weaker peak intensity which is realistic, e.g., for the FLASH
source in Hamburg, $x_\mathrm{quiv} \approx  a$.  
Even more problematic is the fact that the soft-core potential is
harmonic about its minimum with a characteristic frequency
$(Z/a^{3})^{1/2}$ which could become resonant with the VUV laser
frequency.

To avoid these problems we use a different approximative 
potential which has the correct asymptotic Coulomb behavior at large 
distances but lacks an eigenfrequency since it has a non-zero slope 
at $r=0$,
\be{vshape} 
V(r) = - \frac{Z}{r}(1-e^{-r/a})\,. 
\ee
Here, $a$ is chosen in analogy to the potential (\ref{ushape})
discussed above. Note that $U(r{\to}0)=V(r{\to}0)=-Z/a$. 
For rare-gas atoms $a$ is of the order of one:
$a=1.74$ for Xe and $a=1.4$ for Ar.

For the photon frequency used here ($\hbar\omega=\unit{20}{eV}$) there is no
qualitative difference using a U-shape (soft-core) or a V-shape
potential.  The subsequent considerations also do not depend on the
approximate or exact form of the of the Coulomb potential.  Therefore
we will use the generic form $v(i,j)$, which could be either of the three
options:
\be{generic}
v(i,j) =  \left \{ 
  \begin{array}{ll} |\vec r_{i}-\vec r_{j}|^{-1} & 
    \text{exact Coulomb}\\[1mm]
    \left[(\vec r_{i}-\vec  r_{j})^2+a^2)\right]^{-\frac 12} & 
    \text{U-shape} \\[1mm]
    \left(1-e^{-|\vec r_{i}-\vec r_{j}|/a}\right)
    |\vec r_{i}-\vec r_{j}|^{-1} & 
    \text{V-shape}
  \end{array} \right .
\ee
Our numerical examples presented here have been obtained with the 
V-shape potential.

\subsection{Identification of localized and delocalized electrons}
\label{sec:localization}
At photon energies comparable or less than the ionization
potential of a cluster ion ($\hbar\omega\le 30$\,eV), most of
the photo-electrons remain in the cluster, i.e., quasi-free
electrons are produced.  They thermalize quickly, i.\,e., form a
plasma. 

We have to determine which among these quasi-free electrons
travel all over the cluster and visit many ions 
(delocalized quasi-free electrons) and which revolve about a single
ion, that is, are effectively in excited states about an ion (localized
quasi-free electrons).   To do so, we record the revolution
angle $\phi(t)$ of each classical electron about its closest ion as a
function of time.  If the electron $j$ moves for two revolutions
($\phi_{j_\alpha} = 4\pi$) about the same ion $\alpha$ we consider it
as localized, and the period of its motion $T^{j_\alpha}$ is then
given by $2\pi=\phi_{j\alpha}(T^{j_\alpha})$. 

\begin{figure}[tbp]
    \centering
    \includegraphics[width=.9\columnwidth]{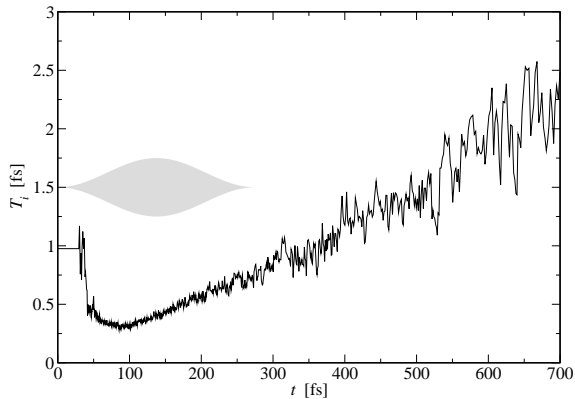}
    \caption{Time intervals $T_i$ of the averaged localized motion 
	of the screening electrons as a function of real time for 
	an Ar$_{147}$ cluster with laser parameter as specified 
	in \sec{sec:run}.}
    \label{fig:intervals}
\end{figure}

The average period $\overline{T}^{j_{\alpha}}$ of all localized classical electrons
sets the time scale for a coarse grained dynamics.
This time scale changes slowly in real time $t$ due to changes in
number and energy of the localized electrons, see \fig{fig:intervals}.
With an initial guess for the first averaged period $T_{0}= 1$\,fs,
we update $T$ after a time $t=T_{i}$ according to the general sequel
\be{avperiod}
\overline{T}{}_{i+1}
=\frac{T_{i}}{n}\sum_{\alpha=1}^{N}\sum_{j_{\alpha}=1}^{n_{\alpha}}
\frac{2\pi}{\phi_{j_\alpha}(T_{i})}\,, \ee where the actual time
interval used is increased by the standard deviation $\sigma_{i+1}$ of
the mean, $T_{i+1} = \overline{T}{}_{i+1}+ \sigma_{i+1}$.

The coarse graining of time through the time intervals $T_{i}$, whose
length is defined by the localized electrons, plays a crucial role for
the description of the entire cluster dynamics.  It provides the
natural time scale to interpolate between the explicit time-dependent
dynamics of the classical electrons and the time-averaged rate
description of the bound ``quantum'' electrons.  Over an interval
$t_{0}<t<t_{0}+T_{i}$ in time all processes involving quantum rates
will be considered within a fixed cluster environment with properties
averaged over the previous interval $[t_0 - T_{i-1}, t_0)$.

\section{Coarse grained photo-ionization into the plasma}
\label{sec:plasma}
We are now prepared to calculate atomic properties in the environment
of other cluster ions and electrons with the understanding that all
these processes are for a specific time interval $T_{i}$ as introduced
in the previous section.  The photo-ionization dipole matrix elements
for many-electron atoms provided within the Hartree-Fock
approximation \cite{co81} allow one to determine the photo cross
section for ionization of individual occupied orbitals to the
continuum, see \sec{sec:photo}.

To apply these cross sections, we have to approximately map the
present situation of a cluster ion surrounded by the localized
electrons and other charged particles (ions and delocalized electrons)
into an effective single ion scenario.  This requires first to
determine the electronic energy of an ion with its localized electrons
and then to construct an energy-equivalent configuration.

\subsection{Electronic energy of the ions including localized
  electrons} \label{sec:enle}
Averaged over  $T_{i}$ we calculate the number 
$n_{\alpha}$ of electrons localized about ion $\alpha$ and their mean 
energy
\be{ealpha}
E^{*}_{\alpha} = E(q_{\alpha}) + \sum_{j=1}^{n_{\alpha}}\left 
(\frac{p_{j}^{2}}{2}-q_{\alpha}v(j,\alpha)\right)+\sum_{j>k=1}^{n_{\alpha}}
v(j,k)\,,
\ee
with $v(j,k)$, the interaction potential \eq{generic} between two
particles of unit charge at positions $\vec r_{j}$ and $\vec r_{k}$,
$q_{\alpha}$, the charge of the ion $\alpha$ and $E(q_{\alpha})$ the
energy of its bound electrons.

\begin{figure}[tbp]
    \centering
    \includegraphics[width=\columnwidth]{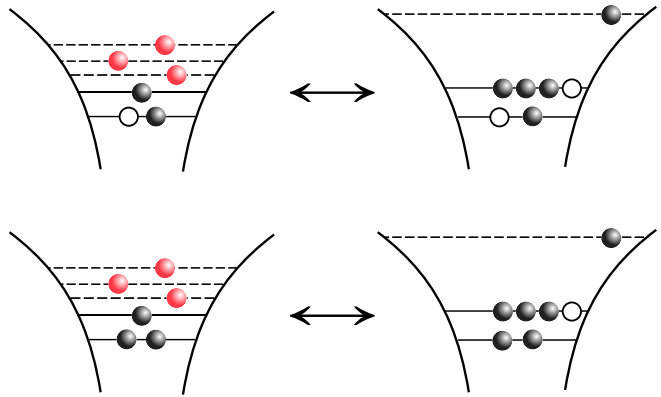}
    \caption{Sketch for the construction of the equivalent electronic
    configuration (right panel) from the actual configuration (left
    panel) of a cluster ion with three localized electrons (red) 
    around a 3s$^{1}$3p$^{1}$ (upper row) and 3s$^{2}$3p$^{1}$ (lower 
    row) configuration.  Holes are shown as open circles.}
    \label{fig:equiv}
\end{figure}

The localized electrons are in excited states of the ion
$\alpha$, as shown in \fig{fig:equiv}. 
Starting from the energy $E^*$ of this actual configuration, we
include them in the photo-ionization process by constructing the
equivalent configuration of ion $\alpha$. 
In this configuration, we relax all localized electrons but one
onto the last occupied orbital of the actual configuration. We
put the remaining electron on a Rydberg orbit, whose energy is given
by the condition that the actual and the equivalent
configuration have the same energy $E^*$. 

\subsection{Ionization potential of the equivalent
  configuration}
\def\C#1#2{C_{#1},#2}
To find the binding energy of electrons in occupied orbitals
in the presence of a Rydberg electron, we assume that neither
the quantum number $n$ nor the angular momentum $\ell$ of the
Rydberg electron change upon release of an electron from a
deeper orbital.  
Then the ionization energy $\Delta E$ is given by
\be{ionpot} \Delta E(\C{q}{n}) = E(\C{q+1}{n}) - E(\C{q}{n})\,, 
\ee
where $E(\C{q}{n})$ is the energy of the valence shell
configuration $C$ with charge $q$ and an additional electron in
a Rydberg orbital $n$. 
We have omitted $\ell$ as an index in (\ref{ionpot}) since the
following discussion does not depend on it.
As an illustrative example we will use $\ell=1$.
\begin{figure}[tbp]
  \centering
  \includegraphics[width=0.8\columnwidth]{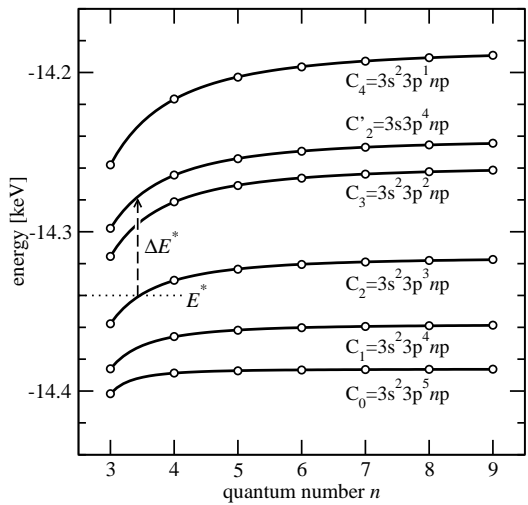}
  \caption{Total energies (circles) for ions with one of
    the electrons in a Rydberg state with quantum number $n$
    and angular momentum $\ell=1$.
    The valence shell configuration $C_q$ is specified for each
    set of energies.
    Fitted curves (thick lines) according to
    Eq.\,(\ref{eq:apen}).
    As an example the energy $E^*$ for an equivalent
    configuration (dotted) as defined in Sect.\,\ref{sec:enle}
    and the corresponding energy $\Delta E^*$ for ionization
    (dashed) are shown.
  }
  \label{fig:energies}
\end{figure}%
The quantum number $n$ can take values from $n_0$ up to
$\infty$, with $n_0$ corresponding to the situation
where the ``Rydberg'' electron is in the lowest possible state;
for Argon and $\ell=1$ it is $n_0=3$.
Figure \ref{fig:energies} provides energies for six
configurations and $n=3\ldots9$.
In five of the configurations the valence shell does not contain
electron holes.  Thus they are solely defined by the charge:
$C_q\equiv q$.  The case $C'_2=$3s3p$^4${}$n$p, also shown in
Fig.\,\ref{fig:energies}, is an example for an exception with a hole
in 3s.  Obviously we have
\be{iopota} E(\C{q}{\infty}) =E(\C{q+1}{n_0})
=E(\C{q}{n_0})+E_\mathrm{ip}(C_q)\,,
\ee
i.\,e.\ the asymptotic energy $n\to\infty$ of a Rydberg series
coincides on one hand with the origin of the Rydberg series of
the next higher charge state, as can be seen in
Fig.\,\ref{fig:energies}. 
On the other hand it is equal to the sum of the ground-state energy 
and the ionization potential $E_\mathrm{ip}$ for an ion with
configuration $C_q$.
For finite values of $n$ we approximate the energies
by a quantum defect formula \cite{brjo03}
\be{eq:apen}
E(\C{q}{n})=E(\C{q}{\infty}) -
\frac 12\left(\frac{Z_{\rm eff}}{n-\mu_{q}}\right)^2\,,
\ee
where in contrast to the usual ionic charge $Z$ we use an
effective one,  
\be{Zeff}
Z_{\rm eff} = 
(n_0-\mu_{q})\left(2E(\C{q}{\infty})-2E(\C{q}{n_0})\right)^{1/2}\,,
\ee
chosen such that the first level ($n_{0}$) of the series agrees with the
exact value while the quantum defect $\mu_{q}$ is fitted.  Hence, 
\eq{eq:apen} is very accurate at intermediate
$n$, where we need it.  When fitting the curves \eq{eq:apen} to the
calculated energies, cf.\ Fig.\,\ref{fig:energies}, we found the
quantum defects $\mu_{q}$ to be almost independent of $q$.  This
allows us  in the calculation of energy differences \eq{ionpot} to
eliminate the term containing $n$ and $\mu$ in \eq{eq:apen}.
We get for a configuration $C_q$ with an initial energy $E^*$
\be{eq:desc}
\Delta E^*(C_q) =
\left[E(\C{q+1}{\infty})-E^{*}\right] 
-b\left[E(\C{q}{\infty})-E^{*}\right]\,,
\ee
with 
\be{eq:defb}
b:=\frac{E_\mathrm{ip}(C_{q+1})}{E_\mathrm{ip}(C_{q})},
\ee
the ratio of the ionization potentials, cf.\
Eq.\,(\ref{iopota}). 
Thus we have obtained an expression for the energy necessary to
ionize an electron from a valence shell in presence of a Rydberg
electron. 
This expression does not depend on the actual quantum numbers
$n$ and $\ell$, but only on the energy $E^*$.

\subsection{The condition for over-barrier inner-ionization}

\begin{figure}[tbp]
  \centering
  \includegraphics[width=0.9\columnwidth]{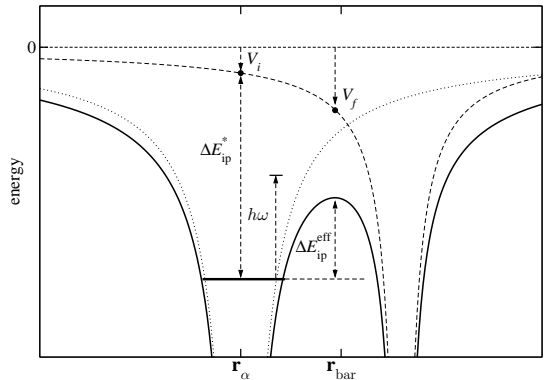}
  \caption{Sketch of the energy balance for the photo-ionization
    of a cluster ion $\alpha$.  
    The cluster potential is represented by a single
    neighbouring ion.  
    The thick full line indicates the full cluster potential,
    the thin long dashed line represents the cluster 
    environment, i.e. the cluster potential without the contribution from
    the ion itself.  The dotted line is the Coulomb potential for
    $q_{\alpha}+1$, i.e. the field of the ion if the electron would be
    ionized.  The interaction of the bound electron with the nucleus is
    represented by $\Delta E^*$.
    The saddle point of $V(\mathbf{r})-(q_{\alpha}+1)v(\alpha,
    \mathbf{r})$ defines the position $\mathbf{r}_{\text{bar}}$
    of the barrier.} 
  \label{fig:barrier}
\end{figure}%
Although we know now with \eq{eq:desc} the ionization potential for a
screened isolated ion, we have to position the ion in the cluster
environment in order to decide, if photo-absorption leads to
photo-excitation within the ion $\alpha$ or to inner-ionization above
the lowest barrier on the way to a neighbouring ion.  The energy balance
for the photo electron which decides between these two options is
\be{total}
E_{i}+V_{i}+\omega=E_{f}+V_{f}
\ee
with local contributions
from the ion to which the electron is bound, (see \fig{fig:barrier})
\be{ef} E_{i} =
-\Delta E^*,\quad
E_{f}=E_{\mathrm{kin}}-(q_{\alpha}+1)v(\alpha,\mathrm{bar})\,, 
\ee
and contributions from the background
of charges in the cluster, 
\be{bg} 
V_{i} = -\sum_{j\ne\alpha}q_{j}v(j,\alpha),\quad
V_{f} = -\sum_{j\ne\alpha}q_{j} v(j,\mathrm{bar})\,, 
\ee
where the index $j$ runs over the delocalized electrons and all ions
but ion $\alpha$.  As introduced earlier, $v(l,m)$ is the
interaction between two Coulomb particles at positions $\vec r_{l}$
and $\vec{r}_m$.  Hence, $V_{i}$ refers to the
potential energy of the electron under consideration and located at
the position of its mother ion $\alpha$ due to interaction with
particles of charge $q_{j}$ at $\vec r_{j}$.  Likewise, $V_{f}$ is the
potential energy of the same electron at the potential barrier $\vec
r_{\rm bar}$ due to the interaction with the same charged particles as
before.  The energy balance \eq{total} is taken with respect to the
location of the ion $\vec r_{\alpha}$ and the location $\vec
r_{\mathrm{bar}}$ of the lowest potential barrier near the ion.
$\Delta E^\mathrm{eff}$ in \fig{fig:barrier} is defined by putting
$E_{\mathrm{kin}}=0$ in \eq{ef}, i.e.,
\be{Ibarrier} \Delta E^\mathrm{eff} = \Delta
E^*-(q_{\alpha}+1)v(\alpha,\mathrm{bar})+V_{f}-V_{i}\,.  
\ee 
\begin{figure}[tbp]
  \centering
    \includegraphics[width=0.9\columnwidth]{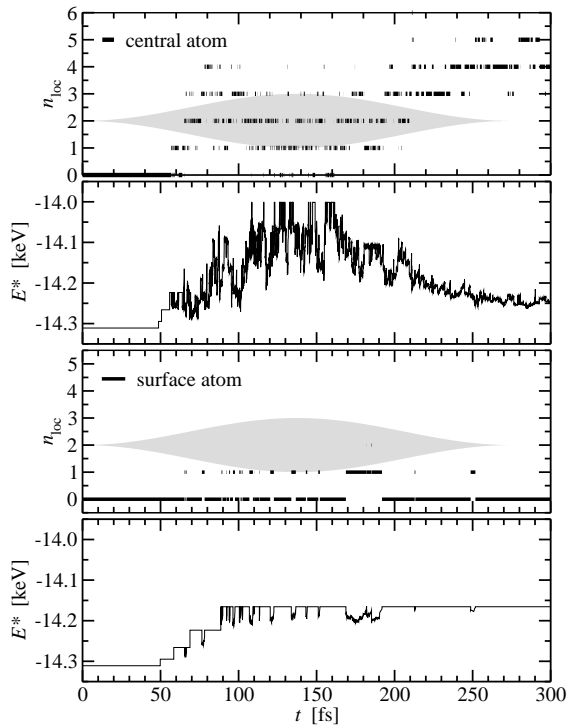}
    \caption{The number of localized quasi-free electrons
      $n_\mathrm{loc}$ and the corresponding energy $E^{*}$
      according to \eq{ealpha} for the central ion and a surface
      ion of an Ar$_{147}$ cluster from the microscopic
      calculation in \sec{sec:run}.}
    \label{fig:fine}
\end{figure}%
Figures \ref{fig:fine} and \ref{fig:coarsegrained} give an overview
of the coarse grained variables during the laser pulse for an atom at
the center and one at the surface of the Ar$_{147}$ cluster.
\Fig{fig:fine} shows first of all an overview of their evolution
during the whole interaction with the laser.  
Starting at the ground state of the neutral Ar, each absorbed
photon leads to a rising step in the total energy. 
Note that for an ion in the cluster the electron has to be
excited only above the lowest barrier. 
Moreover, the energy $E^*$ of the equivalent configuration takes
merely the localized electrons into consideration and not the
newly ionized one. 
Therefore, each ionization event leads to jumps of $E^*$ higher
than the energy of the photon, cf.~\fig{fig:fine}.  

The electron localization is equivalent to a relaxation of the 
system and lowers therefore $E^*$.  
The flat regions observable for both ions correspond to the case
where there are no localized electrons, when the total energy of
the ion is given solely by the bound, ``quantum'' electrons. 
The smaller final charge of the surface atom is a consequence of
the cluster expansion.  
The surface expands much faster than the core, leading to higher
interionic barriers and an early suppression of the
inner-ionization. 
A detailed view of the evolution of the two atoms is shown in
Fig.~\ref{fig:coarsegrained} for the time interval from $t=143$ to
\unit{151}{\femto\second}. 
The coarse graining is symbolized by the full lines, showing the
total energy $E^*_{\alpha}$ of the two atoms averaged over the
time intervals $T_i$, as described in \sec{sec:localization}.
\begin{figure}[tbp]
    \centering
   \includegraphics[width=0.9\columnwidth]{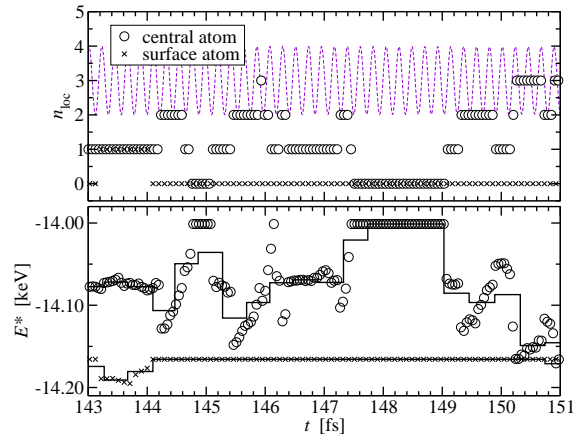}
   \caption{Zoom into \fig{fig:fine} for $143 < t <
     \unit{151}{\femto\second}$. The upper panel shows the
     number of localized electrons ($\circ$ for the central
     atom, where $q=6$ and $\times$ for the surface atom,
     $q=4$). The lower panel shows the total energy $E^*$ of the
     ions and their localized electrons. 
     The full lines show here the average of $E^*$ over the time
     intervals $T_i$, as introduced in \Sec{sec:localization}.}
    \label{fig:coarsegrained}
\end{figure}%

\subsection{The effective cross section for inner photo-ionization}
\label{sec:inner}
Finally, we are in a position to adopt the photo-ionization cross
section as formulated in \sec{sec:photo} for an isolated ion
to the situation of an ion in the cluster.  Here, we take the lowest
potential barrier to a neighboring ion as effective ionization
threshold.  Therefore, the actual cross section as shown in
\fig{fig:photorates} can vary between minimum and maximum possible values
of potential barriers.  The interval is indicated for small photon
energies in \fig{fig:photorates} with the additional dotted line.

The electrons available for photo-ionization are only the tightly 
bound ones from the actual configuration, while the matrix elements 
entering the expression for the cross section \eq{photoabsorb} take into 
account the screening of the electrons as provided by the equivalent 
configuration $\sigma = \sigma(q_{\mathrm{eqv}})$. Hence, the multiplicity 
has to be taken from  the actual configuration $w_{\mathrm{act}}$ to 
arrive at the screened photo cross section
\be{screened}
\sigma^{\mathrm{scr}}(q) = \frac{w_{\mathrm{act}}}{w_{\mathrm{eqv}}} 
\sigma(q_{\mathrm{eqv}})\,.
\ee

\section{Dynamics of Ar$_{147}$ under an intense VUV laser 
pulse}\label{sec:run}

\begin{figure}[b]
  \centering
  \includegraphics[width=0.9\columnwidth]{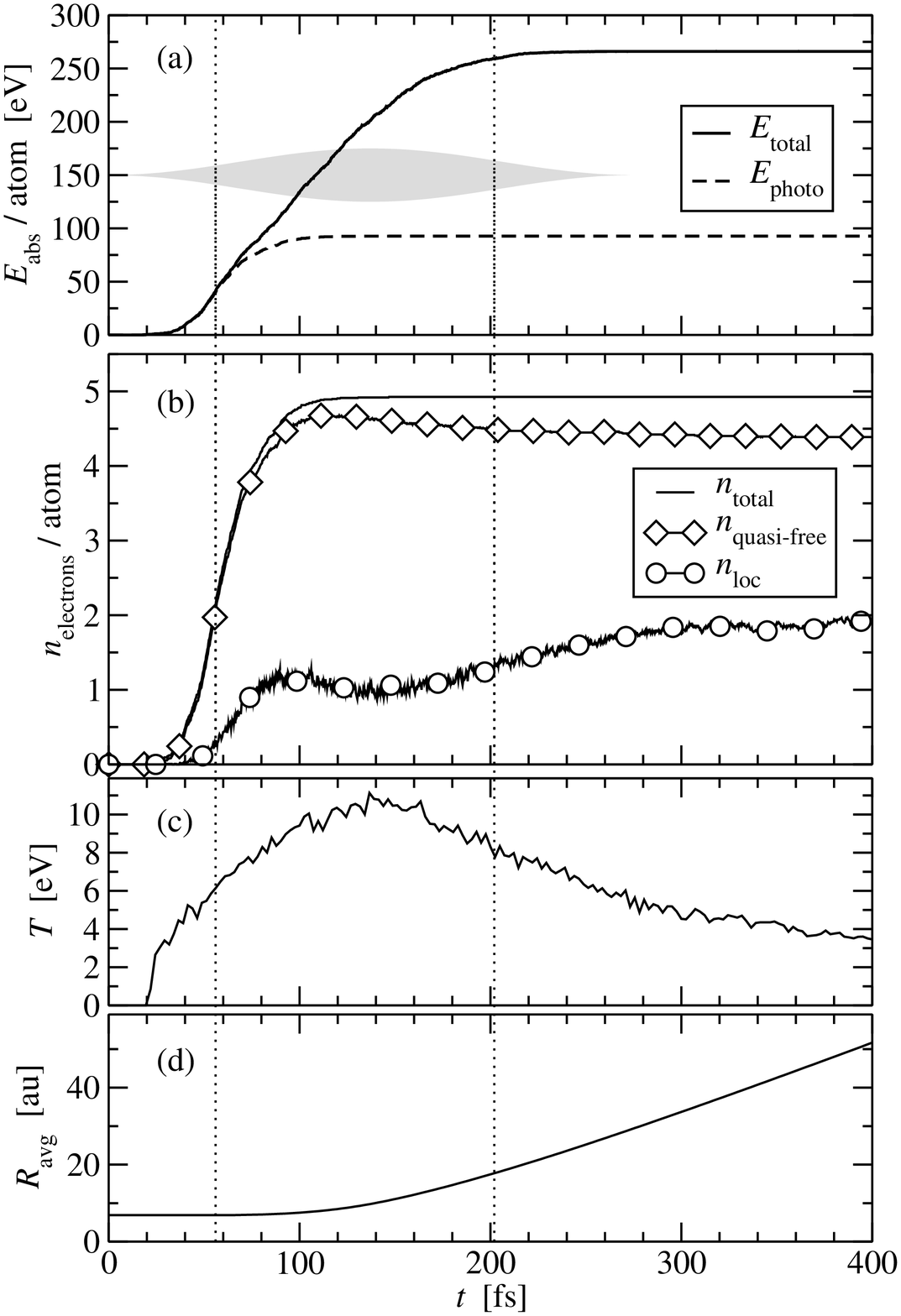}
  \caption{Explosion of Ar$_{147}$ exposed to a \unit{100}{\femto\second} long
    VUV laser pulse with intensity
    $I=\unit{7\times10^{13}}{\watt\per\centi\meter\squared}$ and photon energy
    $\hbar\omega = \unit{20}{\electronvolt}$.  (a)
    full-line: total absorbed energy per atom; dashed-line:
    energy absorbed due to photo-ionization;  (b) full-line:
    total number of ionized electrons per atom; $\diamond$
    quasi-free electrons;  $\circ$ localized electrons; (c)
    temperature of the quasi-free electrons; (d) expansion: 
    average interionic distance, see \eq{ravg}. 
  } 
  \label{fig:Ar147-explosion}
\end{figure}%
We will illustrate the theoretical framework introduced above  with 
the dynamics  of Ar$_{147}$ exposed to a
\unit{100}{\femto\second}  VUV laser pulse with $\hbar\omega =
\unit{20}{\electronvolt}$ (i.\,e.\ wavelength $\lambda=62$\,nm)
and an intensity of
$\unit{7\times10^{13}}{\watt\per\centi\meter\squared}$.  
\Fig{fig:Ar147-explosion} shows the main quantities
characterizing the response of the cluster to the VUV pulse, 
namely the energy absorption, the ionization degree, the
temperature of the plasma and the cluster explosion.
The latter is characterized by the interionic distance
$R_{\text{avg}}$, defined as  
\begin{equation}
R_{\text{avg}} = \sqrt{\frac{1}{N}\sum_{\alpha=1}^N
  \min_{\beta\neq\alpha}(|\mathbf{r}_{\beta} -
  \mathbf{r}_{\alpha}|^2)}. 
\label{ravg}
\end{equation}
In the first part of the pulse, until approximatively
$t=\unit{80}{\femto\second}$, the absorption is dominated by
photo-ionization. 
After that, inverse bremsstrahlung (IBS) sets in and
photo-ionization starts to saturate. 
We can easily disentangle the two contributions to the total
absorbed energy since we treat photo-ionization via rates, see
\sec{sec:photo}, and IBS through classical propagation, see
\sec{sec:clpr}. 

Note that until this time, as can be seen in panel (b) of
\fig{fig:Ar147-explosion}, almost all electrons ($n_\mathrm{total}$)
are trapped ($n_\mathrm{quasi-free}$) inside the cluster due to their
low kinetic energy.  They start to leave the cluster with the onset of
the IBS heating.  This disagrees with the common interpretation that,
when the photon frequency is larger than the first atomic ionization
potential, all cluster atoms are singly ionized by one-photon
absorption.  Rather, the space charge built up even in a relatively
small cluster such as Ar$_{147}$ allows the electrons to become
\emph{quasi}-free only.  Hence, the ``ionization into the cluster'',
as described in \sec{sec:plasma}, starts very early in the pulse.
These electrons thermalize very quickly and obey a
Maxwell-Boltzmann velocity distribution with a temperature $T$
which is shown as function of time in
\fig{fig:Ar147-explosion}c.

\Fig{fig:Ar147-explosion}b also shows the average number of
localized electrons, as they have been introduced in 
\sec{sec:localization}.  
They reach a maximum ($t\approx90$\,fs) when the
photo-ionization becomes unlikely.
The increase of the temperature of the electron plasma
(see \fig{fig:Ar147-explosion}c) from this point on favors
a decrease of the number of localized electrons.
At the point ($t\approx120$\,fs)  where the cluster expansion (see
\fig{fig:Ar147-explosion}d) startes, the temperature of the
electrons plasma decreases, despite of the continuing energy
absorption  and the localization increases again to an average
of two electrons per atom. 
\begin{figure}[htbp]
  \centering
  \includegraphics[width=0.49\columnwidth]{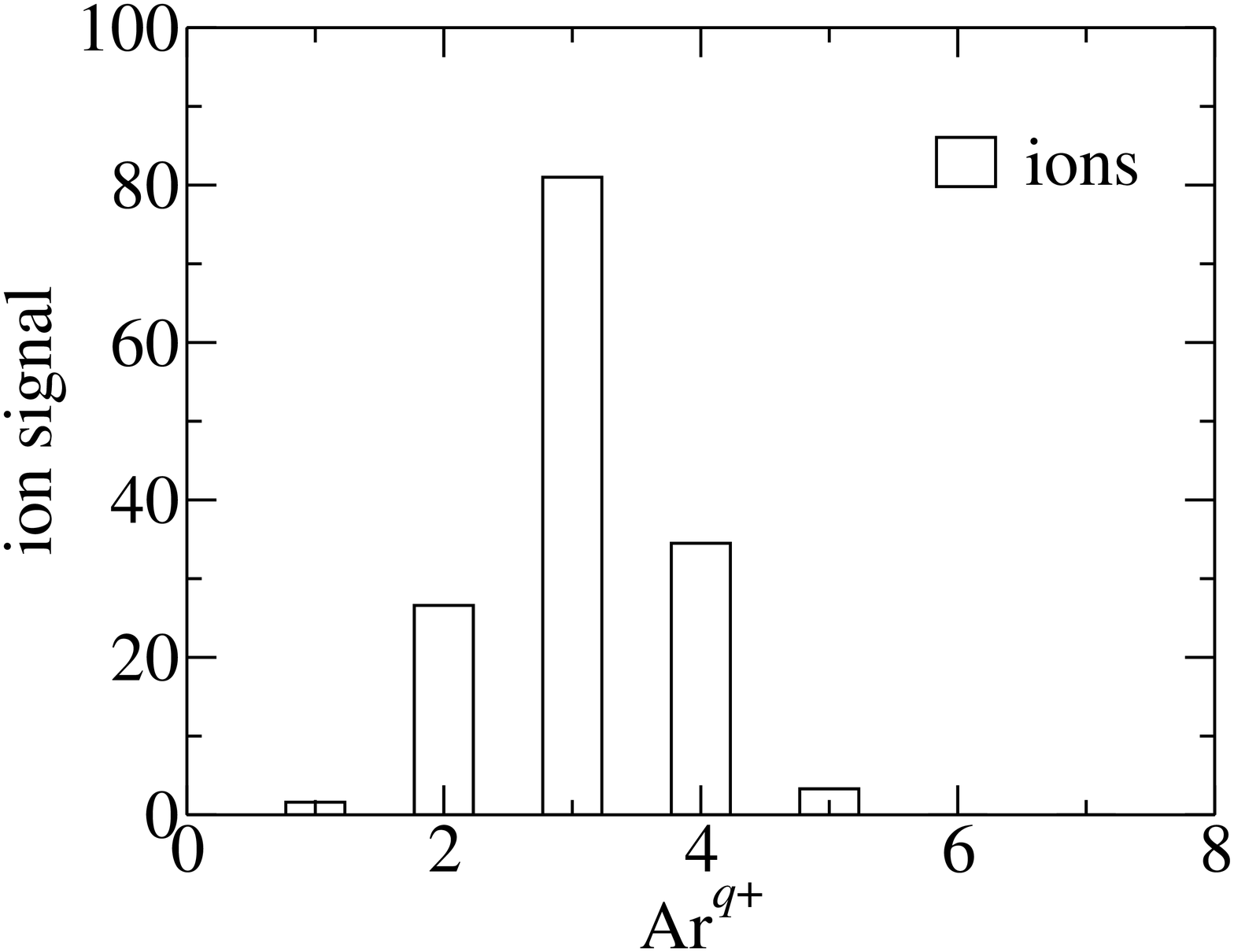}
  \includegraphics[width=0.49\columnwidth]{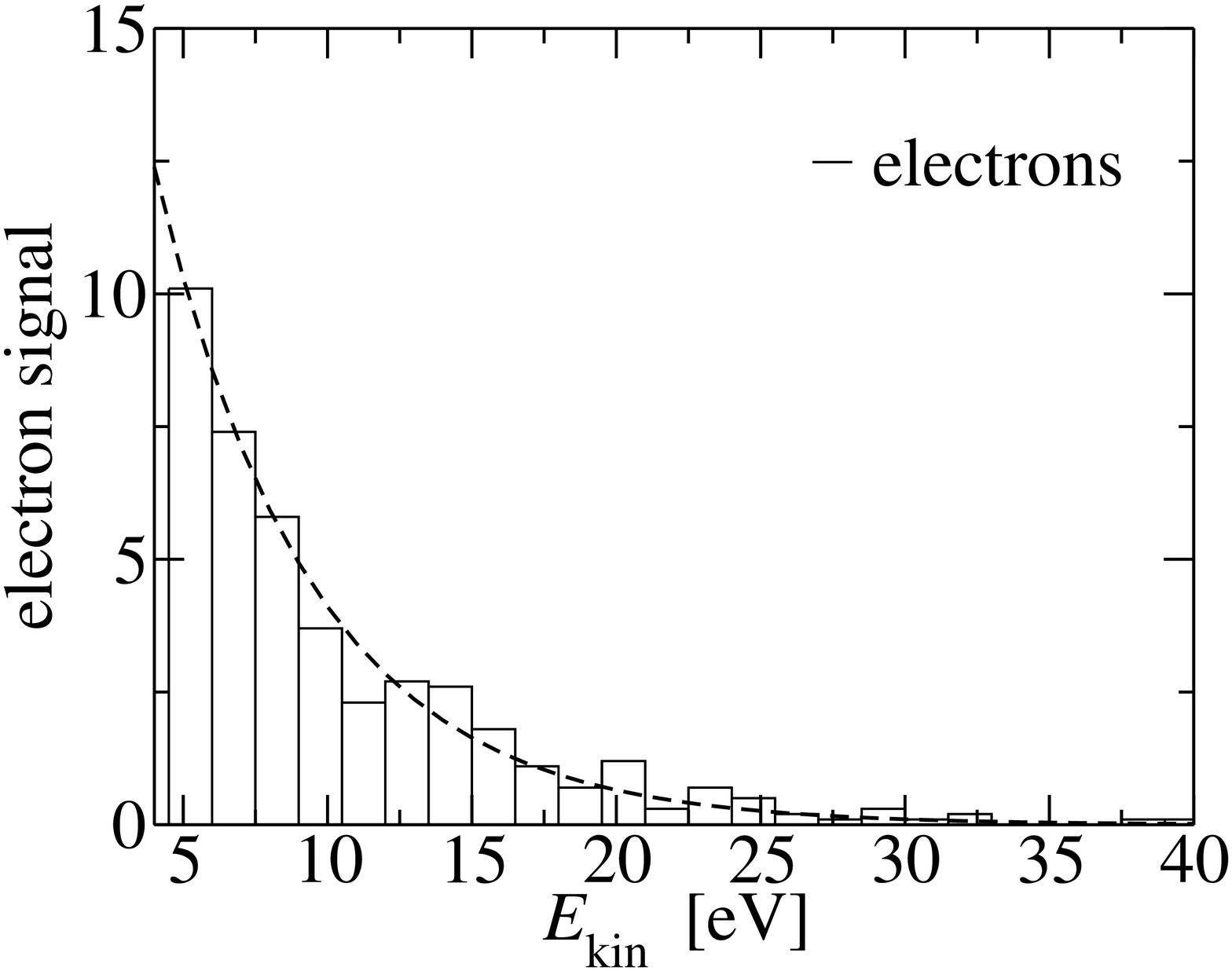}
  \caption{The ion charge distribution (left panel) and the
    kinetic energy distribution of the released electrons
    (right) after the cluster explosion for the same pulse as in
    \fig{fig:Ar147-explosion}.
    The propagation time was $t=\unit{6.4}{\pico\second}$.
    The electron signal can be exponentially fitted by
    $\exp(-E_\mathrm{kin}/E_0)$ with
    $E_0=\unit{5.4}{\electronvolt}$ (dashed line).
}
  \label{fig:Ar147-outcome}
\end{figure}%

This transient complex behavior of electrons and ions in the cluster
may be accessible experimentally soon using attosecond pulse probe
techniques, for a proposal see \cite{gesa+07a}.  For the time being,
we present in \fig{fig:Ar147-outcome} more conventional observables
measurable in the experiment, such as the final charge distribution of
ions and the kinetic-energy distribution of electrons for the dynamics of
Ar$_{147}$.  They have been obtained by propagation up to
$t=\unit{6.4}{\pico\second}$, i.\,e., much longer than shown in
\fig{fig:Ar147-explosion}.  The dominating fragment is Ar$^{3+}$
despite the fact that about five electrons
(\fig{fig:Ar147-explosion}b) per atom have been photo-ionized.  The
 kinetic-energy distribution of the electrons can be fitted with an
exponential decay $\exp(-E_\mathrm{kin}/E_0)$ with
$E_0=\unit{5.4}{\electronvolt}$, thus emphasizing the thermal origin of
the electrons.  The temperature of the plasma has similar values, see
\fig{fig:Ar147-explosion}c.

\section{Summary}\label{sec:sum}
We have described a quantum-classical hybrid approach to follow the
time-dependent dynamics of rare-gas clusters exposed to intense short
laser pulses.  Special attention has been paid to incorporate the
screening of cluster ions by the electron plasma, formed by quasi-free
electrons which have been ionized from their mother ions but cannot
leave the cluster as a whole due to the strong background charge.  The
mean time scale of these localized quasi-free electrons is determined
and provides the link between the microscopic dynamics of all charged
particles and the quantum dynamics of photoionization which is
described by ionization rates adopted to the screening in the cluster
environment.

Hence, this approach is especially well suited to tackle interaction
of clusters with light from VUV and X-FEL laser sources.  As an
illustrative example we have discussed the dynamics of Ar$_{147}$
exposed to a \unit{100}{\femto\second} laser pulse of
$I=\unit{7\times10^{13}}{\watt\per\centi\meter\squared}$ intensity
and $\hbar\omega = \unit{20}{\electronvolt}$ photon energy.

\end{document}